\documentclass[twocolumn,showpacs,aps,preprintnumbers,amsmath,amssymb]{revtex4}

\usepackage{graphicx,subfigure}
\usepackage{dcolumn}
\usepackage{bm}
\usepackage{axodraw4j}

\makeatletter
\newcommand{\fmslash}[2][0mu]{%
  \mathchoice
    {\fmsl@sh\displaystyle{#1}{#2}}%
    {\fmsl@sh\textstyle{#1}{#2}}%
    {\fmsl@sh\scriptstyle{#1}{#2}}%
    {\fmsl@sh\scriptscriptstyle{#1}{#2}}}
\newcommand{\fmsl@sh}[3]{%
  \m@th\ooalign{$\hfil#1\mkern#2/\hfil$\crcr$#1#3$}}
\makeatother

\newcommand{\mpt}{{{\fmslash P}_T}}
\newcommand{\mptvec}{{\vec{\fmslash P}_T}}
\newcommand{\mptvecperp}{{\vec{\fmslash P}_{T_\perp}}}
\newcommand{\mptvecpar}{{\vec{\fmslash P}_{T_\parallel}}}

\begin{document}


\title{Superpartner mass measurements with 1D decomposed $M_{T2}$}

\author{Partha Konar${}^1$, Kyoungchul Kong${}^2$, Konstantin T.~Matchev${}^1$, and Myeonghun Park${}^1$}
\affiliation{${}^1$Physics Department, University of Florida, Gainesville, FL 32611, USA}
\affiliation{${}^2$Theoretical Physics Department, SLAC, Menlo Park, CA 94025}%

\date{19 March, 2010}

\begin{abstract}
We propose a new model-independent technique for mass measurements in 
missing energy events at hadron colliders. We illustrate our method with
the most challenging case of a short, single-step decay chain.
We consider inclusive same-sign chargino pair production in supersymmetry,
followed by leptonic decays to sneutrinos:
$\chi^+\chi^+ \to \ell^+\ell'^+\tilde\nu_\ell\tilde\nu_{\ell'}$
($\chi^-\chi^- \to \ell^-\ell'^-\tilde\nu_\ell^\ast\tilde\nu_{\ell'}^\ast$).
We introduce two one-dimensional decompositions of the Cambridge 
$M_{T2}$ variable:
$M_{T2_\parallel}$ and $M_{T2_\perp}$, on the direction of the upstream 
transverse momentum $\vec{P}_T$ and the direction orthogonal to it, respectively. 
We show that the sneutrino mass $M_c$ can be measured directly by
minimizing the number of events $N(\tilde{M}_c)$ in which $M_{T2}$ 
exceeds a certain threshold, conveniently measured from the endpoint 
$M_{T2_\perp}^{max}(\tilde M_c)$.
\end{abstract}

\maketitle

The Large Hadron Collider (LHC) at CERN has begun its long awaited 
exploration of the TeV scale, where new physics beyond the Standard Model (SM)
may hold the key to our understanding of some very basic questions about
our universe: What is the dark matter? What are the fundamental symmetries 
of Nature? Are there any hidden dimensions of space? A potential 
discovery of a missing energy signal at the LHC may 
relate to all three of these questions, if the missing energy is
due to a stable, neutral, weakly interacting massive
particle in a theory with space-time supersymmetry (SUSY) \cite{Chung:2003fi} 
or extra dimensions \cite{Hooper:2007qk}.

\begin{figure}[t]
\begin{center}
\begin{picture}(180,90)(20,20)
\Text(22,70)[c]{\Black{$p(\bar{p})$}}
\Text(22,40)[c]{\Black{$p(\bar{p})$}}
\SetColor{Gray}
\Line( 40,65)( 55,95)
\DashLine(50,82)( 84,82){2}
\Line( 77,65)( 92,95)
\CBoxc(70,98)(55,15){Red}{Yellow}
\Text( 70,98)[c]{\Black{$U: \vec P_T$}}
\Line(110,65)(125,95)
\DashLine(120,82)(135,82){2}
\Line(130,65)(145,95)
\CBoxc(140,98)(65,15){Red}{Yellow}
\Text(140,98)[c]{\Black{$V_1: \left(m_1,\vec{p}_{1T}\right)$}}
\Line(110,35)(125, 5)
\DashLine(118,25)(133,25){2}
\Line(130,35)(145, 5)
\CBoxc(140,10)(65,15){Red}{Yellow}
\Text(140,10)[c]{\Black{$V_2: \left(m_2,\vec{p}_{2T}\right)$}}
\SetColor{Gray}
\Line(10,65)(80,65)
\Line(10,35)(80,35)
\CBoxc(190,65)(50,15){Red}{Yellow}
\Text(190,65)[c]{\Black{$\left(M_c,\vec{p}_{1T}^{\,\,c}\right)$}}
\CBoxc(190,35)(50,15){Red}{Yellow}
\Text(190,35)[c]{\Black{$\left(M_c,\vec{p}_{2T}^{\,\,c}\right)$}}
\Text(100, 71)[c]{\Red{$P$}}
\Text(100, 28)[c]{\Red{$P$}}
\Text(148, 71)[c]{\Red{$C$}}
\Text(148, 28)[c]{\Red{$C$}}
\SetColor{Red}
\SetWidth{1.5} 
\Line(90,65)(160,65)
\Line(90,35)(160,35)
\Text(102, 49)[c]{\Black{$M_p$}}
\Text(102, 60)[c]{\Black{${\mathbf \uparrow}$}}
\Text(102, 40)[c]{\Black{${\mathbf \downarrow}$}}
\Text(150, 49)[c]{\Black{$M_c$}}
\Text(150, 60)[c]{\Black{${\mathbf \uparrow}$}}
\Text(150, 40)[c]{\Black{${\mathbf \downarrow}$}}
\CBoxc(70,50)(40,50){Blue}{Cyan}
\end{picture}
\end{center}
\caption{\label{fig:event} The generic event topology under consideration.
All particles visible in the detector are clustered into three groups:
upstream objects $U$ with total transverse momentum $\vec{P}_T$,
and two composite visible particles $V_i$, each with invariant mass $m_i$ 
and total transverse momentum $\vec{p}_{iT}$. The transverse momenta of the 
two missing particles are labelled by $\vec{p}_{iT}^{\,\,c}$.}
\end{figure}
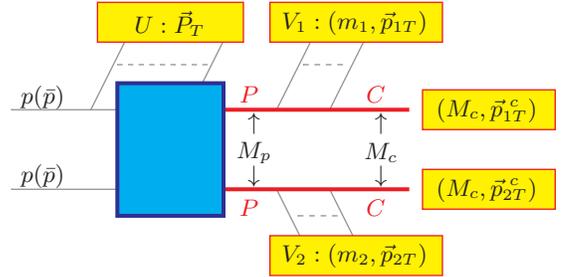

The first order of business after the discovery of a 
missing energy signal at the LHC will be to measure 
the mass of the missing particle and prove that it is
not simply a SM neutrino \cite{Chang:2009dh}. 
This deceptively simple task turned out to be
a notoriously difficult challenge.
The generic topology of a prototypical ``SUSY-like'' missing energy event 
is schematically depicted in Fig.~\ref{fig:event}.
Consider inclusive production of an identical pair of 
new particles $P$ (from now on referred to as ``parents'').
Each parent decays semi-invisibly to a set of SM
particles $V_i$, ($i=1,2$), which are visible in the detector, 
and a dark matter particle $C$ (from now on referred to as the ``child'') 
which escapes detection.
In general, the parent pair may be accompanied by
a number of additional ``upstream'' objects $U$ (typically jets)
with total transverse momentum $\vec{P}_T$. They
may originate from various sources such as initial state radiation 
or decays of even heavier particles up the decay chain.
We shall not be interested in the exact details of the 
physics responsible for $U$, adopting a fully inclusive approach 
to the production of the parents $P$. Given this general setup, 
the goal is to determine {\em independently} the mass $M_p$ of the parent
and the mass $M_c$ of the child. 

In the past, several approaches to this problem have been proposed,
e.g.~invariant mass endpoint measurements \cite{imass}
or exact reconstruction of the missing particle momenta $\vec{p}_{iT}^{\, c}$
\cite{exactreco}. Unfortunately, they only apply to sufficiently long 
decay chains, where the visible particles in $V_i$ arise from 
a sequence of at least three 2-body decays \cite{Burns:2008va}.
In the simplest example of a short, single-step decay chain, each $V_i$ consists 
of a single SM particle of fixed mass $m_i$, and neither of these 
two approaches will work. 
One must then resort to methods 
based on the Cambridge $M_{T2}$ variable \cite{approxreco,kink,Matchev:2009fh} 
or the related Sheffield $M_{CT}$ variable \cite{Tovey:2008ui,Matchev:2009ad}.
Unfortunately, in order to apply those techniques, one must 
work with a subset of events within a relatively narrow fixed 
$P_T$ range, incurring some loss in statistics. 

In this Letter we propose a new method which 
uses the full data set, with no such loss in statistics.
Our method is based on the ``subsystem'' variant \cite{Matchev:2009fh}
of the original $M_{T2}$ variable \cite{approxreco}.
For any given event, one can construct the transverse mass 
$M_{iT}$ of each parent $P$:
\begin{equation}
M_{iT}^2\equiv m_i^2+M_c^2+2(E_{iT} E_{iT}^c-\vec{p}_{iT}\cdot\vec{p}^{\, c}_{iT})\, ,
\label{MTparent}
\end{equation}
where 
\begin{equation}
E_{iT}   \equiv \sqrt{m_i^2+|\vec p_{iT}|^2}, \quad
E_{iT}^c \equiv \sqrt{M_c^2+|\vec p_{iT}^{\,c}|^2},
\label{E}
\end{equation}
is the transverse energy of the visible particle $V_i$ 
and child particle $C$ in each branch of Fig.~\ref{fig:event},
correspondingly. The individual momenta $\vec p_{iT}^{\,c}$ 
of the missing child particles $C$ are unknown, but
they are constrained by the measured missing transverse 
momentum $\mptvec$ in the event:
\begin{equation}
\vec p_{1T}^{\,c} + \vec p_{2T}^{\,c} = \mptvec
\equiv - \vec{P}_T - \vec p_{1T}- \vec p_{2T}. 
\label{PTmiss}
\end{equation}
For the true values of the missing momenta $\vec p_{iT}^{\,c}$, 
each transverse mass in (\ref{MTparent}) is bounded from 
above by the true parent mass $M_P$. This fact can be used 
in a rather ingenious way to define the Cambridge $M_{T2}$ 
variable \cite{approxreco}. One takes the larger of the two
quantities in (\ref{MTparent}) and minimizes it over
all possible partitions of the unknown children momenta 
$\vec p_{iT}^{\,c}$, subject to the constraint (\ref{PTmiss}):
\begin{equation}
M_{T2}
\equiv 
\min_{\vec p_{1T}^{\,c} + \vec p_{2T}^{\,c} = \mptvec }
\left\{\max\left\{M_{1T},M_{2T}\right\} \right\}\ .
\label{eq:mt2def}
\end{equation}
For a given $P_T$, the endpoint $M_{T2}^{max}$ of this distribution 
gives the parent mass $\tilde M_p$ as a function of the input 
trial child mass $\tilde M_c$:
\begin{equation}
\tilde M_p(\tilde M_c, P_T) \equiv M_{T2}^{max}(\tilde M_c, P_T)\, .
\label{Mptildedef}
\end{equation}
This property 
provides one relation among the two unknown 
masses $M_p$ and $M_c$ \cite{approxreco}.


Here we propose to obtain a second relation by using 
the property that
the function $\tilde M_p(\tilde M_c, P_T)$
is independent of $P_T$ at the true child mass $M_c$:
\begin{equation}
\tilde M_p(M_c, P_T+\Delta P_T) - \tilde M_p(M_c, P_T) = 0,
~\forall\, \Delta P_T,
\label{ptindependence}
\end{equation}
which we can rewrite more informatively as
\begin{equation}
\tilde M_p(\tilde M_c, P_T) - \tilde M_p(\tilde M_c, 0) \ge 0\, ,
\label{ptdependence}
\end{equation}
with equality being achieved only for $\tilde M_c=M_c$.
Eq.~(\ref{ptdependence}) implies that,
for any given $\tilde M_c$, there will always be a certain
number of events whose $M_{T2}$ values will exceed the 
reference value $\tilde M_p(\tilde M_c,0)$, unless the trial mass 
$\tilde M_c$ happens to coincide with the true child mass $M_c$.
In order to quantify this effect, we define the function
\begin{equation}
N(\tilde M_c) \equiv 
\sum_{\textrm{all events}} 
H \left( M_{T2}-\tilde M_p(\tilde M_c,0)\right),
\label{Ndef}
\end{equation}
where $H(x)$ is the Heaviside step function. 
From the definition of
$N(\tilde M_c)$ it is clear that it is minimized at $\tilde M_c=M_c$,
where in theory we would expect
\begin{equation}
N_{min}\equiv \min\{N(\tilde M_c)\}=N(M_c)=0\, .
\end{equation}
In reality, the value of $N_{min}$ will be lifted from 0, due to
finite particle width effects, detector resolution, etc. 
Nevertheless we expect that the {\em location} of the 
$N(\tilde M_c)$ minimum will still be at $\tilde M_c=M_c$,
allowing a direct measurement of the child mass $M_c$:
\begin{equation}
M_c = \left\{ \tilde M_c \, | \, N(\tilde M_c)=N_{min}\right\},
\label{Mcmeasurement}
\end{equation}
which is our first main result.
Once the child mass $M_c$ is found from (\ref{Mcmeasurement}),
the true parent mass $M_p$ is obtained as usual from (\ref{Mptildedef})
as $M_p = \tilde M_p (M_c,P_T)$.

At this point it is not clear whether we have gained anything
statistics-wise, since the reference quantity $\tilde M_p(\tilde M_c,0)$ 
appearing in the definition (\ref{Ndef}) has to be
measured at a fixed $P_T=0$ anyway. Our second main result 
in this paper is that $\tilde M_p(\tilde M_c,0)$ can in fact 
be measured from the full data set with no loss in statistics
as follows.

Let us introduce one-dimensional (1D)
decompositions of $M_{T2}$ onto the two special directions 
defined by the upstream momentum vector $\vec{P}_T$.
Following Ref.~\cite{Matchev:2009ad}, first project the visible 
transverse momenta $\vec{p}_{iT}$ of Fig.~\ref{fig:event}
onto the $\vec{P}_T$ direction ($T_\parallel$) and its orthogonal direction 
($T_\perp$):
\begin{eqnarray}
\vec p_{iT_\parallel} &\equiv& \frac{1}{P_T^2}\left(\vec{p}_{iT}\cdot\vec{P}_T\right)\vec{P}_T ,
\label{pitpar} \\
\vec p_{iT_\perp} &\equiv& \vec{p}_{iT}-\vec p_{iT_\parallel}
= \frac{1}{P_T^2} \vec{P}_T \times \left(\vec{p}_{iT}\times \vec{P}_T\right),
\label{pitperp}
\end{eqnarray}
and similarly for the two transverse momenta $\vec{p}^{\, c}_{iT}$
of the children and for $\mptvec$. Now consider the corresponding 1D decompositions
of the transverse parent masses (\ref{MTparent})
\begin{eqnarray}
M_{iT_\parallel}^{2} &\equiv&
m_i^2 + \tilde M_c^2 + 2\left(E_{iT_\parallel} E_{iT_\parallel}^c 
- \vec{p}_{iT_\parallel}\cdot\vec{p}_{iT_\parallel}^{\, c} \right), \nonumber
\label{MTparallel}
\\
M_{iT_\perp}^{2}  &\equiv& 
m_i^2 + \tilde M_c^2 + 2\left(E_{iT_\perp} E_{iT_\perp}^c
- \vec{p}_{iT_\perp}\cdot\vec{p}_{iT_\perp}^{\, c} \right),\nonumber
\label{MTperp}
\end{eqnarray}
in terms of the 1D projected analogues of (\ref{E})
\begin{eqnarray}
E_{iT_\parallel} &\equiv& \sqrt{m_i^2+|\vec p_{iT_\parallel}|^2}, \quad
E_{iT_\perp}     \equiv \sqrt{m_i^2+|\vec p_{iT_\perp}|^2}, \nonumber \\
E_{iT_\parallel}^c &\equiv& \sqrt{\tilde M_c^2+|\vec{p}_{iT_\parallel}^{\, c}|^2 }, \quad
E_{iT_\perp}^c \equiv \sqrt{\tilde M_c^2+|\vec{p}_{iT_\perp}^{\, c}|^2 }.
\nonumber
\end{eqnarray}
Now we define 1D $M_{T2}$ decompositions
in complete analogy with the standard $M_{T2}$ definition 
(\ref{eq:mt2def}):
\begin{eqnarray}
M_{T2_\parallel} 
&\equiv& 
\min_{\vec p_{1T_\parallel}^{\,c} + \vec p_{2T_\parallel}^{\,c} = \mptvecpar }
\left\{\max\left\{M_{1T_\parallel},M_{2T_\parallel}\right\} \right\} ,
\label{eq:mt2parallel} \\
M_{T2_\perp}    
&\equiv& 
\min_{\vec p_{1T_\perp}^{\,c} + \vec p_{2T_\perp}^{\,c} = \mptvecperp}
\left\{\max\left\{M_{1T_\perp},M_{2T_\perp}\right\} \right\}.
\label{eq:mt2perp}
\end{eqnarray}
These decompositions
are extremely useful. For once, the 1D variables 
(\ref{eq:mt2parallel},\ref{eq:mt2perp}) can be calculated 
via simple analytic expressions as shown below. In contrast,
a general formula for the original $M_{T2}$ variable (\ref{eq:mt2def})
in the presence of arbitrary $P_T$ 
is unknown and one still has to compute $M_{T2}$
numerically \cite{Cheng:2008hk}. 
More importantly, $M_{T2_\perp}$
allows us to measure the reference quantity 
$\tilde M_p(\tilde M_c,0)$ in (\ref{Ndef})
from the full data set, using events with {\em any} 
value of $P_T$.

To understand the basic idea, 
it is sufficient to consider the simplest, yet most
challenging case of a single step decay chain. Let
$V_i$ be a single, 
(approximately) massless SM particle: $m_1=m_2=0$.
(The discussion for the massive case proceeds analogously.)
In what follows, for illustration we shall use the
same-sign dilepton channel in supersymmetry, where
each $V_i$ is a lepton resulting from a chargino decay 
to a sneutrino \cite{Matchev:2009fh}. The charginos themselves
are produced indirectly in the decays of squarks and gluinos.
For concreteness we shall use a SUSY spectrum given by the LM6 CMS 
study point \cite{Ball:2007zza}.
At point LM6, the chargino (sneutrino) mass is
$M_p=305.3$ GeV ($M_c=275.7$ GeV), and the rest of the 
SUSY mass spectrum can be found in \cite{Ball:2007zza}.
In our simulations we use the PYTHIA event generator \cite{Sjostrand:2006za}
and the PGS detector simulation program \cite{PGS}.

\begin{figure}[t]
\includegraphics[width=7.0cm]{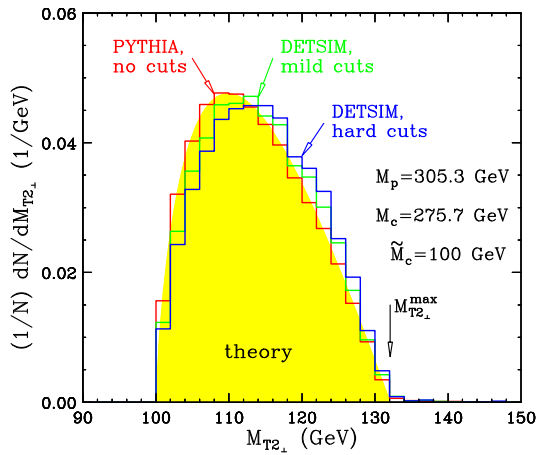}
\caption{\label{fig:perp} The unit-normalized $M_{T2_\perp}$ distribution
(\ref{dNbar}) for the same-sign dilepton channel in a SUSY model with 
LM6 CMS mass spectrum and a choice of test mass $\tilde M_c = 100$ GeV.
The yellow shaded distribution shows the theoretically predicted shape (\ref{dNbar}),
matching very well the parton level result from PYTHIA with no cuts (red histogram).
The green (blue) histogram is the corresponding result after
PGS detector simulation with mild
(hard) cuts as explained in the text.
The endpoint expected from eq.~(\ref{eq:mt2perpmax}) is $132.1\,$GeV
and is marked with the vertical arrow.}
\end{figure}

The variable $M_{T2_\perp}$ has several unique properties.
Eventwise, it can be calculated analytically as
\begin{eqnarray}
M_{T2_\perp}&=&\sqrt{A_{T_\perp}}+\sqrt{A_{T_\perp}+\tilde{M}_c^2}~~, 
\label{eq:mt2perpEvent}  \\
A_{T_\perp}&\equiv& \frac{1}{2}
\left(|\vec{p}_{1T_\perp}| | \vec{p}_{2T_\perp}|
      +\vec{p}_{1T_\perp}\cdot \vec{p}_{2T_\perp}\right).
\nonumber
\end{eqnarray}
The endpoint of the $M_{T2_\perp}$
distribution is given by
\begin{equation}
M_{T2_\perp}^{max}(\tilde M_c)=\mu+\sqrt{\mu^2+\tilde{M}_c^2},
\label{eq:mt2perpmax}
\end{equation}
in terms of the parameter $\mu$ introduced in \cite{Burns:2008va}
\begin{equation}
\mu\equiv \frac{M_p}{2}\left(1-\frac{M_c^2}{M_p^2}\right).
\end{equation}
Eq.~(\ref{eq:mt2perpmax}) reveals perhaps the most important 
feature of the $M_{T2_\perp}$ variable: its endpoint is 
{\em independent} of the upstream $P_T$ and can thus 
be measured with the {\em whole} data sample.
We can even predict analytically the 
{\em shape} of the (unit-normalized) 
differential $M_{T2_\perp}$ distribution
\begin{equation}
\frac{\mathrm{d} N}{\mathrm{d}M_{T2_\perp} } =
N_{0_\perp}\, \delta(M_{T2_\perp}-\tilde M_c)
+
\left(1-N_{0_\perp}\right)
\frac{\mathrm{d} \bar{N}}{\mathrm{d}M_{T2_\perp}},
\label{dNperp}
\end{equation}
where
$N_{0_\perp}$ is the fraction of events in the lowest $\tilde M_c$ bin
$M_{T2_\perp}=\tilde M_c$, while the shape of the remaining (unit-normalized) 
$M_{T2_\perp}$ distribution is given by (see Fig.~\ref{fig:perp})
\begin{equation}
\frac{\mathrm{d}\bar{N}}{\mathrm{d}M_{T2_\perp} } = 
\frac{M_{T2_\perp}^4-\tilde{M}_c^4}{\mu^2\,M_{T2_\perp}^3}
  \ln\left( \frac{2\mu\, M_{T2_\perp} }{M_{T2_\perp}^2-\tilde{M}_c^2 }\right).
\label{dNbar}
\end{equation}
Notice that this shape
does not depend on any unknown kinematic parameters,
such as the unknown center-of-mass energy 
or longitudinal momentum of the initial hard scattering. 
It is also insensitive to spin correlation effects, whenever the 
upstream momentum results from production and/or decay 
processes involving scalar particles (e.g.~squarks) or 
vectorlike couplings (e.g.~the QCD gauge coupling). 
It is even independent of 
the actual value of the upstream momentum 
$P_T$. Thus 
we are not restricted to a particular $P_T$ range and can use 
the {\em whole} event sample in the $M_{T2_\perp}$ analysis.
For any choice of $\tilde M_c$ (in Fig.~\ref{fig:perp} we used
$\tilde M_c = 100$ GeV), eq.~(\ref{dNbar}) is a one-parameter curve
which can be fitted to the data to obtain the parameter $\mu$ and 
from there the $M_{T2_\perp}$ endpoint (\ref{eq:mt2perpmax}).

\begin{figure}[t]
\includegraphics[width=7.0cm]{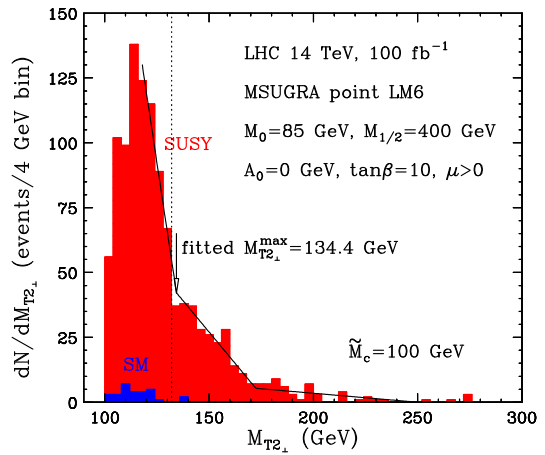}
\caption{\label{fig:pgsdist} Observable $M_{T2_\perp}$ distribution
after hard cuts for 100 fb${}^{-1}$ of LHC data. The total stacked distribution 
consists of the SUSY signal (red) and the SM background (blue).
The solid line is the result of a simple linear fit, revealing
endpoints at 134.4 GeV and 172.4 GeV.}
\end{figure}

As always, there are practical limitations to the use of 
such shape fitting. 
First, the shape (\ref{dNbar}) is modified in the presence of ``mild'' cuts, 
which are required for lepton identification in PGS 
(green histogram in Fig.~\ref{fig:perp}), 
and more importantly, for the discovery of the same-sign 
dilepton SUSY signal over the SM backgrounds. 
To ensure discovery, we use ``hard'' cuts 
as follows 
\cite{Ball:2007zza,Pakhotin:2006wx}: exactly
two isolated leptons with $p_T>10$ GeV, 
at least three jets with $p_T>(175,130,55)$ GeV,  
$\mpt > 200$ GeV and a veto on tau jets.
With those cuts, in the dimuon channel alone, the 
remaining SM background cross-section is
dominated by $t\bar{t}$ and is just 0.15 fb, while
the SUSY signal is 14 fb, leading to a
$22\sigma$ discovery with just $10\ {\rm fb}^{-1}$ 
of data \cite{Ball:2007zza,Pakhotin:2006wx}.
The distortion of the $M_{T2_\perp}$ shape with these
hard offline cuts is illustrated by the blue 
(rightmost) histogram in Fig.~\ref{fig:perp}.
The actual $M_{T2_\perp}$ distribution which we expect to
observe with $100\ {\rm fb}^{-1}$ of data, is shown 
in Fig.~\ref{fig:pgsdist} and is comprised of a relatively small 
SM background component (blue) and a dominant SUSY signal 
component (red). In spite of the presence of a sizable
SUSY combinatorial background, the $M_{T2_\perp}$ endpoint
expected from Fig.~\ref{fig:perp} is clearly visible
and its location from a simple linear fit is obtained as
134.4 GeV, which is very close to the nominal value of
132.1 GeV. (Interestingly, the data reveals a second endpoint at
172.4 GeV, which is due to events in which one 
chargino decays through a charged slepton: 
$\tilde\chi^\pm_1\to\tilde \ell_L^\pm\to \tilde\chi^0_1$
\cite{Matchev:2009fh}.
Its nominal value is 169.2 GeV.)

Our final key observation is that 
\begin{equation}
\tilde M_p(\tilde M_c,0) = M_{T2}^{max}(\tilde M_c,0) = M_{T2_\perp}^{max}(\tilde M_c),
\end{equation}
which allows to rewrite 
the function $N(\tilde M_c)$ of eq.~(\ref{Ndef}) as
\begin{equation}
N(\tilde M_c) \equiv 
\sum_{\textrm{all events}} 
H \left( M_{T2}-M_{T2_\perp}^{max}(\tilde M_c)\right).
\label{Ndef2}
\end{equation}

\begin{figure}[t!]
\includegraphics[width=7.0cm]{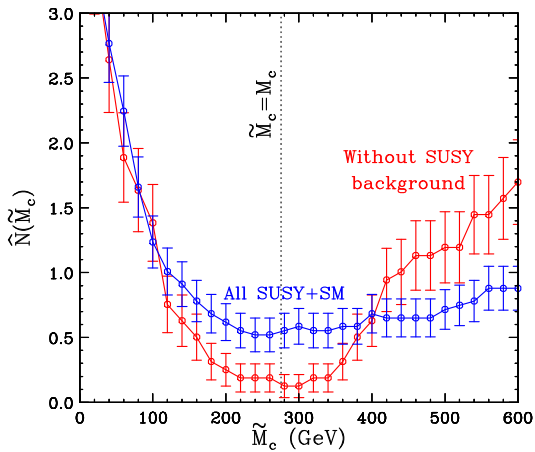}
\caption{\label{fig:N} The function $\hat{N}(\tilde M_c)$
defined in (\ref{Ndef3}). The blue (red)
set of measurements are with (without) 
SUSY combinatorial background.
The error bars shown are purely statistical.
}
\end{figure}

The $M_{T2_\perp}$ analysis just described 
allows a very precise measurement of the 
benchmark quantity $M_{T2_\perp}^{max}(\tilde M_c)$
appearing in (\ref{Ndef2}), so that the function
$N(\tilde M_c)$ itself can be reliably reconstructed, 
using {\em the whole} event sample all the way 
throughout the analysis, without any loss in statistics.
We show our result in Fig.~\ref{fig:N}, where for convenience we
unit-normalize the function $N(\tilde M_c)$ as
\begin{equation}
\hat{N}(\tilde M_c) = N(\tilde M_c)/\langle N(\tilde M_c)\rangle,
\label{Ndef3}
\end{equation}
where the averaging is performed over the plotted range of $\tilde M_c$.
As expected, the function $\hat{N}(\tilde M_c)$ 
exhibits a minimum in the vicinity of the true
sneutrino mass $\tilde M_c=M_c=275.7$ GeV.
Ignoring the SUSY combinatorial background, 
this measurement (red data points) is quite precise, 
at the level of a few percent. In order to reduce the combinatorial 
background, we select events with 
$\tilde M_c < M_{T2_\perp} < M^{max}_{T2_\perp}$
and veto very hard\footnote{The 
measured value of $M_{T2_\perp}^{max}$ in Fig.~\ref{fig:pgsdist}
already implies that the mass splitting $M_p-M_c$ is on the order of 
30 GeV, resulting in a rather soft lepton $p_T$ spectrum.} leptons
with $p_T>60$ GeV. The resulting $M_c$ measurement 
(blue data points) is at the level of $10\%$.
This precision is clearly sufficient 
to exclude SM neutrinos as the source of the missing energy,
hinting at a potential dark matter discovery at the LHC. 

In conclusion, we summarize the novel features and advantages of our method
in comparison to previous $M_{T2}$-based proposals in the literature
\cite{kink,Matchev:2009fh}. First, we make crucial use of 
property (\ref{ptindependence}), which allowed us to 
measure {\em directly} the missing particle mass $M_c$ as 
in eq.~(\ref{Mcmeasurement}). Second, both the benchmark quantity
$M_{T2_\perp}^{max}(\tilde M_c)$ entering eq.~(\ref{Ndef2})
as well as the the function $N(\tilde M_c)$ itself
can be measured using the whole available data sample
at {\em any} $P_T$. 
To the extent that the definition of $M_{T2_\perp}$ relies only
on the {\em direction} and not the magnitude of the upstream $\vec{P}_T$,
our method is insensitive to the jet energy scale error \cite{Matchev:2009ad}.
We have also provided exact analytical formulas for the computation
of the 1D decomposed $M_{T2}$ variables\footnote{The corresponding 
analytical results for $M_{T2_\parallel}$ can be found in the first 
version of this paper, which is available on the hep-ph archive.} 
and the shape (\ref{dNbar}) of the $M_{T2_\perp}$ distribution.

{\em Acknowledgments.} We thank L.~Pape for useful comments.
This work is supported in part by
US Department of Energy grants DE-FG02-97ER41029 and DE-AC02-76SF00515.

\end{document}